\documentclass[twocolumn, showpacs]{revtex4}
\usepackage{graphics}
\usepackage{amsmath}
\usepackage{mathrsfs}
\usepackage{amsfonts}
\usepackage{overpic}
\usepackage{hyperref}
\usepackage{rotating}
\newcommand{\ket}[1]{|#1\rangle}
\newcommand{\bra}[1]{\langle #1|}

\begin{document}
\title{Geometric Phase and Quantum Phase Transition : Two-Band Model}
\author{H. T. Cui}
\email{cuiht@aynu.edu.cn} \affiliation{Department of Physics, Anyang
Normal University, Anyang 455000, China}
\author{Jie Yi}
\affiliation{School of Physical Science and Technology, Heilongjiang
University, Harbin 150018, China }
\date{\today}
\begin{abstract}
The connection between the geometric phase and quantum phase
transition has been discussed extensively in the two-band model. By
introducing the twist operator, the geometric phase can be defined
by calculating its ground-state expectation value. In contrast to
the previous numerical examinations, our discussion presents an
exact calculation for the determination of the geometric phase.
Through two representative examples, our calculation shows the
intimate connection between the geometric phase and phase
transition: different behaviors of the geometric phase can be
identified in this paper, which are directly related to the energy
gap above the ground state.
\end{abstract}
\pacs{75.10.Pq, 03.65.Vf, 05.30.Pr, 42.50.Vk} \maketitle

\section{introduction}
Recently,  quantum phase transition \cite{sachdev} has received
great attention due to its intimate correlation to the fundamental
principles of quantum mechanics, especially to the concept of
quantum entanglement \cite{preskill, osterloh, wu, vidal} ( see
Ref.\cite{afov07} for a comprehensive review ). In general the
quantum phase transition happens when the degeneracy of the ground
states occurs, which cannot be characterized completely by the
pattern of symmetry broken ( order parameters of some kind ).
Instead the universal quantum order or topological order is needed
for the description of properties of the ground state in many-body
systems \cite{wen}. Recently the connection between the geometric
phase of the ground state and the quantum criticality has been
displayed in spin-chain systems by displaying the singularity of the
geometric phase closed to the critical points \cite{carollo, zhu}.
Moreover in Ref \cite{zhu} the author showed that the scaling
behavior of the geometric phase of the ground state near the
critical points can also display the universal class of the phase
transitions. Many works have been devoted to this interesting issue
\cite{cui, rhp07, Zanardi, pv} ( or see Ref.\cite{zhu08} for a
review ).

Although  great progresses has been made in the understanding of
quantum phase transition from the fundamental principles of quantum
mechanics, an important question is not yet resolved: whether there
exists a universal way to characterize the different phases and
their boundaries. More specifically, could the geometric phase of
the ground state in many-body systems serve this purpose? This
conjecture is natural since the quantum phase transition generally
emerges from the degeneracy of the ground state in many-body
systems. Geometric phase as a measurement of the curvature of the
Hilbert space, could mark the fantastic changes of the ground state
when degeneracy happens. However the critical point is how to obtain
the geometric phase of the ground state. The earlier method is to
impose a local rotation about some special orientations, as has been
done in \cite{carollo, zhu, cui}. In our own viewpoints, the
connection built by this method is fragile; the geometric phase is
trivial when the system is symmetrical about this rotation.

Recently the differential information-geometry analysis of quantum
fidelity in many-body systems displayed the intimate correlation
between quantum phase transition and the singularity of fidelity
between the states across the transition point \cite{Zanardi}. In
these papers the quantum geometric tensor, which is intimately
related to the degeneracy of the ground state, was introduced for
the determination of the fidelity. As shown in these papers, the
imaginary part of this tensor actually described the curvature
two-form whose holonomy is the Berry phase, and the degeneracy of
the ground state would induce the singularity of fidelity
\cite{Zanardi}. However, in our opinion, it seems not transparent to
directly define the geometric phase in this coupling-parameter space
since a cyclic evolution may be difficult to construct. Moreover,
the explicit expression for the ground state is necessary for the
construction of this tensor, which in general is difficult.
Furthermore it is also unknown in the case where the degeneracy of
the ground state is broken. In Ref. \cite{rhp07}, the Bargmann
phase, a generalization of the Berry phase, has also been
constructed for detecting the phase transition in many-body systems.
However the connection between degeneracy of the ground state and
the Bargmann phase was unclear in this case since there was a lack
of simple interpretational underpinnings for the Bargmann phase in
terms of physical adiabatic processes \cite{rhp07}.

With respect to the points stated above, it is urgent to find a
popular way for construction of geometric phase in many-body
systems. For this purpose, a nonlocal operator -the twist operator-
is introduced to obtain the geometric phase of the ground state in
this paper. Our calculation shows that the geometric phase, decided
by the ground-state expectation value of the twist operator, can
serve as the quantity to distinguish different phases and boundaries
between them, as shown below. The general form for the twist
operator can be written as in the lattice systems,
\begin{equation}\label{tw}
\eta= \exp(\frac{2\pi i}{N}\sum_x xn_x),
\end{equation}
in which $N$ is the number of lattice sites, $x$ denotes the
coordinates of lattice sites and  $n_x$ is generally related to the
physical quantities located at site $x$, such as the total spin or
charge, or particle number at site $x$ and so on.  The twist
operator was first introduced by Lieb, Schultz and Mattis for the
proof of gapless excitation in one-dimensional spin-1/2
chains\cite{lsm}. Then Resta pointed out that its ground state
expectation was direct to the Berry-phase theory of polarization in
strongly correlated electron systems \cite{resta}. Moreover Aligia
found that the ground-state expectation value of the twist operator
Eq. \eqref{tw} allows one to discriminate conducting from
nonconducting phase in the extended quantum systems \cite{aligia}.
The vanishing of the ground-state expectation value, i.e. $\eta=0$,
has been shown the ability to detect the boundaries of different
valence-bond-solid phases in spin chains \cite{naka}. However these
studies were implemented in some special examples and a general
discussion was absent. Moreover since the previous calculations were
numerical or approximate, the details of $\eta$ adjacent to the
phase transition points are unclear. Hence it is of great interest
to find the exact expression for $\eta$, even for special cases. Our
paper serves this purpose, and the exact results can be found for a
special case.

It is of great interest to note that the twist operator $\eta$
actually creates a wave-like excitation since it rotates all the
particles with a relative angle between the neighboring lattices
$2\pi /N$\cite{lsm}. Under the large $N$ limit the ground state has
an adiabatic variation, and its ground-state expectation value is
exactly a geometric factor, of which an argument is the geometric
phase. Applying $\eta$ to the unique ground state, one obtains a
low-lying excited state. The important quantity is the overlap
between the ground state and this excited state, i.e.
\begin{equation}\label{z}
z = \bra{g}\eta\ket{g},
\end{equation}
in which $\ket{g}$ denotes the many-body ground state. In general
$z$ is a complex number and its argument is just the geometric
phase, determined by
\begin{equation}\label{gp}
\gamma_g= \text{Arg}[z]=i
\int_0^{2\pi}d\phi\bra{g(\phi)}\partial_{\phi}\ket{g(\phi)},
\end{equation}
in which $\ket{g(\phi)}=\exp(\frac{i\phi}{N}\sum_x
x\hat{n}_x)\ket{g}$. Since $\gamma_g$ in fact came from the
continued deformation of the boundary condition of systems
\cite{aligia} and then slightly related to the symmetry of the
Hamiltonian, this construction of the geometric phase is more
popular than the previous method. An important character is that
$\gamma_g$ is related to the correlation functions for the ground
state, and numerical evaluation could be implemented efficiently
\cite{resta, aligia, naka}.

It is an immediate speculation that $z$ and $\gamma_g$ may be
singular near the critical points, where the degeneracy of the
ground state happens and the macroscopic properties of the system
have fantastic changes. However, our findings are more subtle; the
exact calculations show an unexpected ability for $\gamma_g$ or $z$
to distinguish the different phases in many-body systems; one case
is that $z$ tends to be zero and then $\gamma_g$ is ill-defined when
one approaches the phase transition points, in which the degeneracy
of the ground state happens. The other is that $\gamma_g$ has
different values for different phases and displays the singularity
at the transition points, where no degeneracy happens. The physical
reason, as shown in the following discussions, is directly related
to the energy gap above the ground state.

The paper is organized as follows.  In Sec.II, the exact expression
of $z$ and the geometric phase $\gamma_g$ are presented in the
two-band model for a special case, in which the ground state is the
filled Fermi sea. In Sec. III, two representative examples are
provided for the demonstration of this connection. One is the
$D$-dimensional free-fermion model, in which there are quantum phase
transitions originated from the ground-state-energy degeneracy. The
other example is the Su-Schrieffer-Heeger ( SSH ) model, in which
the energy gap is non-vanishing at the quantum phase transition
points and a topological order, defined by the geometric phase
$\gamma_g$, provides a clear description of the phase diagram for
this model.

\section{two-band model}
Consider the one-dimensional (1D) translational invariant
Hamiltonian with two bands separated by a finite gap \cite{ryu,
ryu2},
\begin{equation}\label{hx}
H = \sum_{x, x'}\mathbf{c}_x^{\dagger}\mathscr{H}_{x,
x'}\mathbf{c}_{x'},
\end{equation}
in which $\mathbf{c}_x^{(\dagger)}= (c_+,
c_-)_x^{\textrm{T}(\dagger)}$ defines a pair of fermion annihilation
(creation) operators for each site $x, x'= 1,2, \dots, N$ and the
form of $\mathbf{c}_x$ is decided completely by the Hamiltonian.
$\mathscr{H}_{x, x'}$ is a $2\times2$ matrix and its elements can be
determined by the hermiticity of Eq. \eqref{hx},
\begin{equation}
\mathscr{H}_{x,x'}=\left(\begin{array}{cc}A_+ & B \\C & A_-
\end{array}\right)_{x,x'}
\end{equation}
in which $A_{\pm, xx'}=A_{\pm,x'x}^*$ and $B_{x,x'}=C^*_{x',x}$.
Although our discussion is restricted to 1D system in this section,
it should point out that this situation can be easily generalized to
higher dimension systems.

Without the loss of generality, it is conventional for the
translational invariant system to impose the periodic boundary
condition. In spite of the simplicity, the Hamiltonian Eq.
\eqref{hx} has a wide range of applications, such as the Bogoliubov-
de Gennes Hamiltonian in superconductivity, graphite systems
\cite{ryu}. Applying the Fourier transformation and considering  the
periodic boundary condition, $\mathbf{c}_x=1/\sqrt{N}\sum_k
e^{ikx}\mathbf{c}_k$ in which $k=2\pi n/N$ with $n=1,2,\dots, N$.
Then the Hamiltonian in the momentum space can be written generally
as $H = \sum_k \mathbf{c}_k^{\dagger}\mathscr{H}(k)\mathbf{c}_k$. If
one introduces the four-vector $R_{\mu}(k)(\mu=0,x,y,z)$, then the
Hamiltonian can be rewritten as,
\begin{equation}\label{hk}
H=\sum_{\mu}\sum_k
\mathbf{c}_k^{\dagger}R_{\mu}(k)\sigma_{\mu}\mathbf{c}_k,
\end{equation}
in which $\sigma_0$ is a $2\times 2$ unit matrix and
$\sigma_{i}(i=x,y,z)$ is the Pauli operators. Obviously Eq.
\eqref{hk} can be diagonalized by finding the eigenvectors
$\nu_{\pm}$ of
$\sum_{i}R_i(k)\sigma_i=\mathbf{R}(k)\cdot\mathbf{\sigma}$, in which
the vector $\mathbf{R}(k)$ is similar to the Bloch vector for the
density operator in the $2\times2$ Hilbert space \cite{footnote},
and furthermore should also satisfy the relation $(R_{x}(k),
R_{y}(k))\neq(0, 0)$ since in this case one has a trivial geometric
phase \cite{berry},
\begin{equation}\label{nu}
\nu_{\pm}= \frac{1}{\sqrt{2R(k)(R(k)\mp
R_{z}(k))}}\left(\begin{array}{c}R_{x}(k) - i R_{y}(k)\\ \pm
R(k)-R_{z}(k)\end{array}\right)
\end{equation}
in which $R(k)=|\mathbf{R}(k)|$ and the corresponding eigenvalues
are $E_{\pm}=R_0(k) \pm R_{z}(k)$. Obviously there is a finite gap
between the two bands since $E_+>E_-$. The ground state is defined
as the filled Fermi sea  $\ket{g}=\prod_k \beta_{-,
k}^{\dagger}\ket{0}_k$, in which $\beta_{-,
k}^{\dagger}=\mathbf{c}_k^{\dagger}\nu_- $ and $\ket{0}_k$ is the
vacuum state of $c_{\pm k}$.

Now it is time to determine the geometric phase Eq. \eqref{gp},
given by the ground-state expectation value of the twist operator
Eq. \eqref{z}. In this  model the twist operator can be expressed
explicitly as
\begin{equation}\label{eta}
\eta=\exp(\frac{2\pi i}{N}\sum_x x
\mathbf{c}_x^{\dagger}\mathbf{c}_x),
\end{equation}
in which $n_x=\mathbf{c}_x^{\dagger}\mathbf{c}_x=c_+^{\dagger}c_+ +
c_-^{\dagger}c_-$. Seemingly  $\mathbf{c}_x^{\dagger}\mathbf{c}_x$
could define the particle number at the site $x$, however the
physical meanings for it may be different for different systems,
dependent on one's interests; for spin systems it may denote the
total spin at site $x$, and for electron systems it may also denote
the total charge number at site $x$. The geometric phases in the
both situations have been defined respectively as the spin Berry
phase and the charge Berry phase, which have extensive applications
in determining the phase diagram in strongly correlated electron
systems \cite{yama}.

It is a crucial step to determine $z$. First with the periodic
boundary condition, one can rewrite $\eta$ in the moment space
\begin{equation}\label{twk}
\eta=\exp(-\frac{2\pi}{N}\sum_k \mathbf{c}_k^{\dagger}\partial_k
\mathbf{c}_k).
\end{equation}
Now, introduce the new fermion operators
\begin{equation}\label{beta}
\beta_{\pm, k}=\nu_{\pm}^{\dagger}\mathbf{c}_k,
\end{equation}
in which $\nu_{\pm}$ is defined in Eq. \eqref{nu} and both of
$\beta_{\pm, k}$ satisfy the anti-commutative relation.  Then the
ground state is defined as the filled Fermi sea
$\ket{g}=\prod_k\beta^{\dagger}_{-, k}\ket{0}_k$. Substitute Eq.
\eqref{beta} into Eq. \eqref{twk}
\begin{equation}\label{twb}
\eta=\prod_k
\exp[-\frac{2\pi}{N}(\mathbf{\beta}_k^{\dagger}\mathcal{M}\mathbf{\beta}_k+\beta_{-,k}^{\dagger}\partial_k\beta_{-,k}+
\beta_{+,k}^{\dagger}\partial_k\beta_{+,k})],
\end{equation}
in which $\mathbf{\beta}_k=(\beta_{-, k}, \beta_{+, k})^T$ and
\begin{equation}
\mathcal{M}=\left(\begin{array}{cc}K'&K\\-K^*&K'^*\end{array}\right),
\end{equation}
where
\begin{eqnarray}
\cos\theta_k&=&\frac{R_z(k)}{R(k)},
\gamma_k=\arctan\frac{R_y(k)}{R_x(k)}\nonumber\\
K'&=&- i \sin^2\frac{\theta_k}{2}\partial_k\gamma_k \nonumber\\
K&=&
\frac{e^{i\gamma_k}}{2}(\partial_k\theta_k+i\sin\theta_k\partial_k\gamma_k)
\end{eqnarray}

The last two terms in Eq. \eqref{twb} precludes the further exact
calculations. An important case is that if $R_{y}(k)=-R_{y}(-k)$,
one then can properly choose $\nu_{\pm}$ so that the new Fermi
operator $\beta_{\pm, k}$ can be converted into each other by
exchanging $k\leftrightarrow -k$. Then the terms in Eq. \eqref{twb},
$\beta_{-,k}^{\dagger}\partial_k\beta_{-,k}, \beta_{+,
k}^{\dagger}\partial_k\beta_{+, k}$ can cancel each other. An exact
result in this special case can be obtained for $z$
\begin{equation}\label{zk}
z=\bra{g}\eta\ket{g}=\prod_k [1-
\frac{|K|^2}{C_+^2}(e^{\frac{2\pi}{N}\lambda_+}-1)-
\frac{|K|^2}{C_-^2}(e^{\frac{2\pi}{N}\lambda_-}-1)]
\end{equation}
in which
\begin{eqnarray}
\lambda_{\pm}&=&\pm i\sqrt{|K'|^2+|K|^2}\nonumber\\
C_{\pm}^2&=&-|K|^2+(\lambda_{\pm}-K')^2
\end{eqnarray}
and from this formula, the geometric phase $\gamma_g$ can also be
obtained exactly.

The exact determining of $z$ provides the ability to detect the
distinguished behaviors of $\gamma_g$ or $z$ near the phase
transition points. Although the exact results can be obtained only
for this special case, some different connections between the
geometric phase $\gamma_g$ or $z$ and quantum phase transitions are
disclosed, as shown in the following calculations.

\section{exemplifications}
In this section two representative models are presented to display
the distinguished characteristics of $\gamma_g$ or $z$ for the
determination of the phase diagram in many-body systems. One is the
$D$-dimensional free-fermion model, in which the quantum phase
transition is originated from the degeneracy of the ground-state
energy \cite{li06}. The other is the Su-Schrieffer-Heeger (SSH)
model, in which the quantum phase transition happens with the
non-vanishing energy gap above the ground state \cite{ryu}. It is
obvious that these examples include two important cases of quantum
phase transition; one is the degeneracy of the ground-state energy,
whereas not for the other. Moreover the both models are exactly
solvable.

\begin{figure}
\includegraphics[width=8cm]{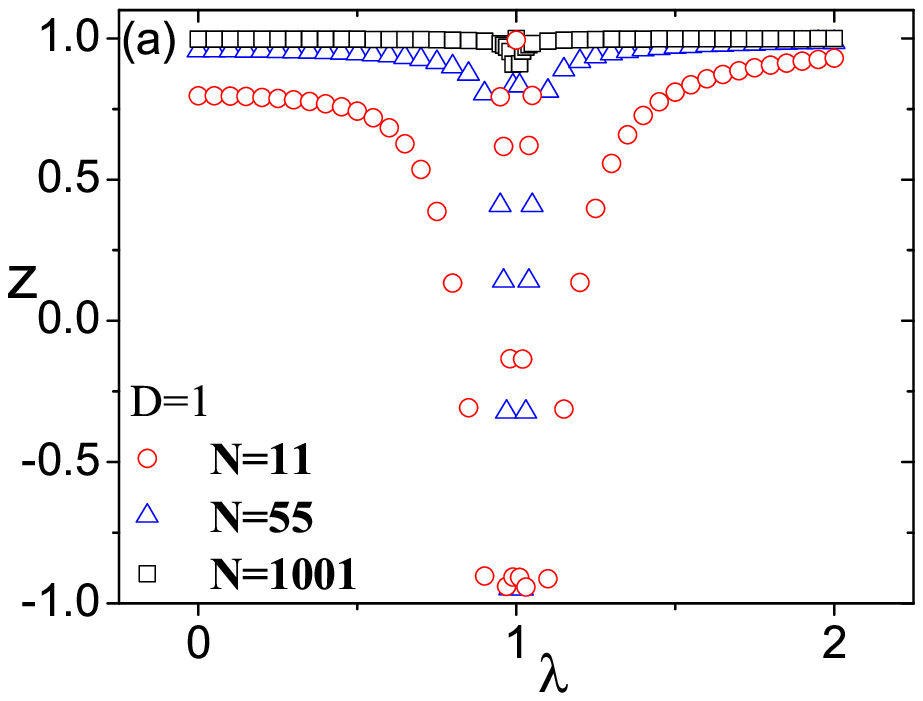}
\includegraphics[width=8cm]{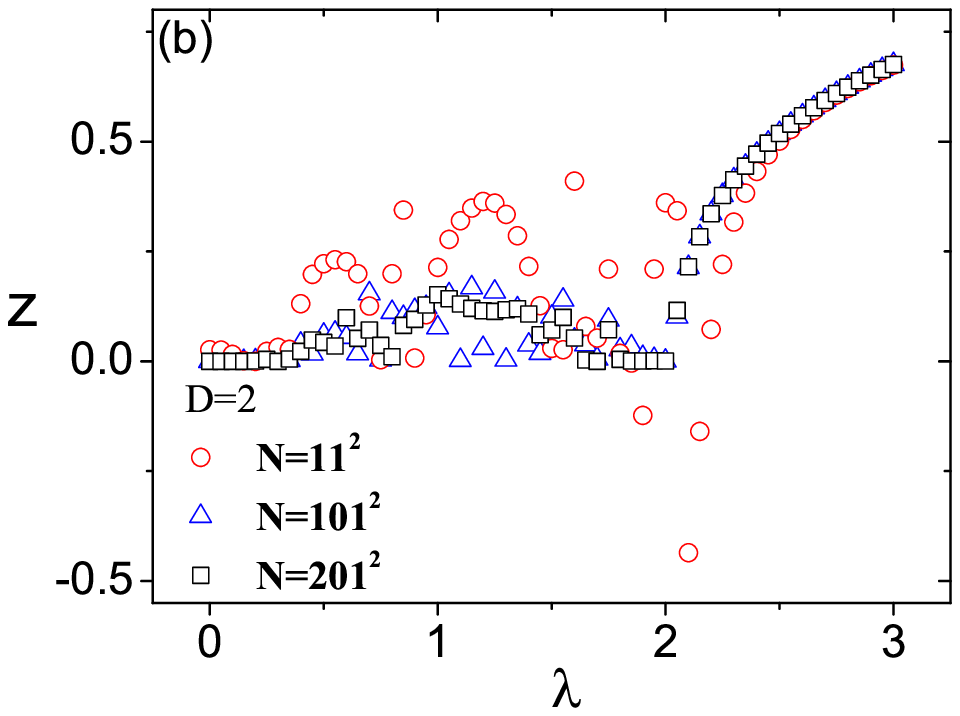}
\includegraphics[width=8cm]{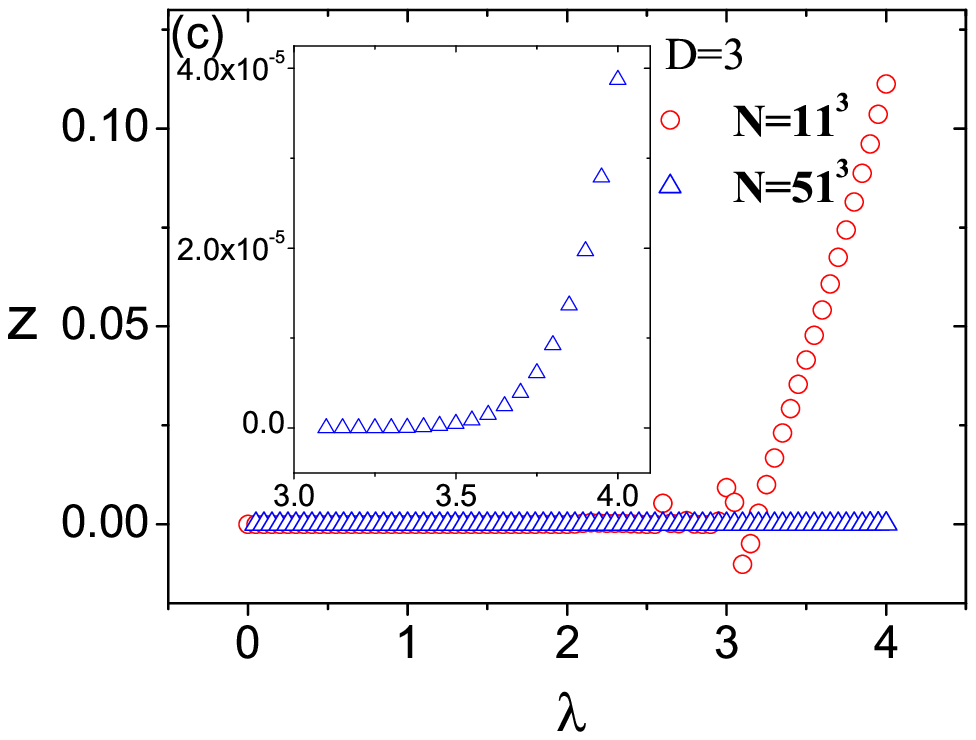}
\caption{\label{ff}( Color online ) The ground-state expectation
value of twist operator $z$ for the $D$-dimensional free-fermion
model vs the parameter $\lambda$. We have chosen $\gamma=1$ for this
plot. The plotting in the inset of (c) shows the details of
$\lambda>3$.}
\end{figure}

\subsection{$D$-dimensional free-fermion model}

The Hamiltonian is read as
\begin{equation}\label{hd}
H=\sum_{\langle {\mathbf{i,j}} \rangle}[c_{\mathbf{i}}^\dagger
c_{\mathbf{j}}-\gamma(c_{\mathbf{i}}^\dagger
c_{\mathbf{j}}^\dagger+h.c.)]-2\lambda\sum_{\mathbf{i}}
c_{\mathbf{i}}^\dagger c_{\mathbf{i}},
\end{equation}
in which $\langle \mathbf{i,j} \rangle$ denotes the nearest-neighbor
lattice sites and $c_{\mathbf{i}}$ is the fermion operator. This
Hamiltonian, first introduced in Ref.\cite{li06}, depicts the
hopping and pairing between nearest-neighbor lattice sites, in which
$\lambda$ is the chemical potential and $\gamma$ is the pairing
potential. Eq.\eqref{hd} could be considered as a $D$-dimensional
generalization of one-dimensional spin-$1/2$ $XY$ model. However for
the $D>1$ case, this model shows some novel phase characteristics
\cite{li06}.

Eq. \eqref{hd} can be resolved exactly by transforming into moment
space with periodic boundary condition.  Wiht the help of the
Bogoliubov transformation \cite{li06}, one has
\begin{equation}
H=\sum_{\mathbf{k}}2\Lambda_{\mathbf{k}}\eta_{\mathbf{k}}^\dagger\eta_{\mathbf{k}}+const.
\end{equation}
in which
$\Lambda_{\mathbf{k}}=\sqrt{t_{\mathbf{k}}^2+\Delta_{\mathbf{k}}^2}$,
$t_{\mathbf{k}}=\sum_{\alpha=1}^D\cos k_\alpha-\lambda$ and
$\Delta_{\mathbf{k}}=\gamma\sum_{\alpha=1}^D\sin k_{\alpha}$. The
phase diagram can be determined based on the gapless excitation
$\Lambda_k=0$ \cite{li06}. For $D=1$, which corresponds to the one
dimensional spin-$1/2$ $XY$ model, the energy gap above the ground
state is non-vanishing except at $\lambda_c=1$ for $\gamma\neq0$,
where a second-order quantum phase transition occurs. For $\gamma=0$
the energy of the ground state is degenerate in the region
$|\lambda|\le1$ and the transition occurs at $\lambda=\pm1$. When
$D=2$, the phases diagram should be identified with respect to two
different situations; for $\gamma=0$, the degeneracy of the ground
state occurs when $\lambda\in[0, 2]$, whereas the gap above the
ground state is non-vanishing for $\lambda>2$. However for
$\gamma\neq0$ three different phases can be identified as
$\lambda=0$, $\lambda\in(0, 2]$ and $\lambda>2$. The first two
phases correspond to case that the energy gap for the ground state
vanishes, whereas not for $\lambda>2$. One should note that
$\gamma=0$ means a well-defined Fermi surface with $k_x=k_y\pm\pi$,
whose symmetry is lowered by the presence of $\gamma$ terms. For
$D=3$ two phases can be identified as $\lambda\in[0,3]$ with the
vanishing energy gap above the ground state and $\lambda>3$ with a
non-vanishing energy gap above ground state. In a word the critical
points can be identified as $\lambda_c=D (D=1,2,3)$ for any
anisotropy of $\gamma$, and $\lambda=0$ for $D=2$ with $\gamma>0$.

Defining the fermion-pair operator $\mathbf{c}_{\mathbf{i}}=(c,
c^{\dagger})^T_{\mathbf{i}}$ and transforming the system into moment
space, one then obtain
\begin{eqnarray}\label{hdk}
H=\sum_{\mathbf{k}}\mathbf{c}^{\dagger}_{\mathbf{k}}\left(\begin{array}{cc}\lambda-\sum_{\alpha}\cos
k_{\alpha}&-i\gamma\sum_{\alpha}\sin
k_{\alpha}\\i\gamma\sum_{\alpha}\sin
k_{\alpha}&-\lambda+\sum_{\alpha}\cos
k_{\alpha}\end{array}\right)\mathbf{c}_{\mathbf{k}}
\end{eqnarray}
It is worth  noting $\mathbf{R(k)}=(0, \gamma\sum_{\alpha}\sin
k_{\alpha}, \lambda-\sum_{\alpha}\cos k_{\alpha})$ and
$R\mathbf{(k)}=\Lambda_{\mathbf{k}}$. $R_{y}(\mathbf{k})$ is
obviously satisfied the requirement
$R_{y}(\mathbf{k})=-R_{y}(-\mathbf{k})$. Substituted into Eq.
\eqref{zk}, one obtain
\begin{eqnarray}
\gamma_g=\text{Arg} z_{\mu}=\text{Arg}\prod_{\alpha=1}^
D\prod_{k_{\alpha}, k_{\mu}}\cos\frac{2\pi}{N}|K_{\mu}|,
\end{eqnarray}
in which $\mu=1\dots D$ and
\begin{widetext}
\begin{eqnarray}\label{kff}
K_{\mu}=-\frac{i\gamma(\lambda-\sum_{\alpha}\cos k_{\alpha})[\cos
k_{\mu}(\lambda-\sum_{\alpha}\cos k_{\alpha})-\sin
k_{\mu}\sum_{\alpha}\sin k_{\alpha}]}{2[\gamma^2(\sum_{\alpha}\sin
k_{\alpha})^2+(\lambda-\sum_{\alpha}\cos k_{\alpha})^2]^{3/2}}.
\end{eqnarray}
\end{widetext}

The schematic drawing of $z$ for $D=1, 2, 3$ have been presented in
Figs. \ref{ff}, in which we have chosen $\gamma=1$ for
specification. Some different characters can be found in the figures
\cite{note}.

\begin{figure}
\includegraphics[width=5cm]{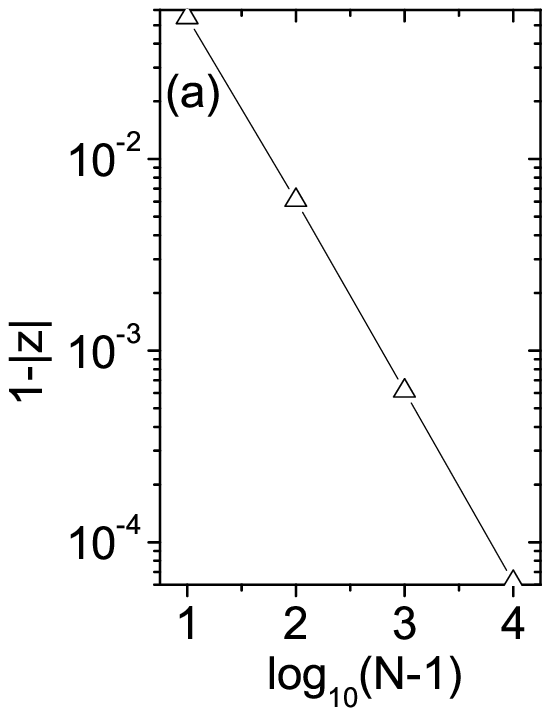}\hspace{-1cm}
\includegraphics[bbllx=1, bblly=15, bburx=165, bbury=218,width=4cm]{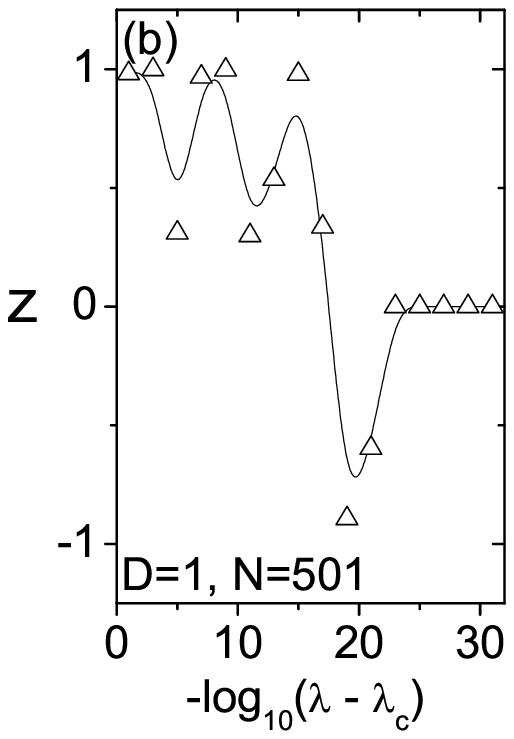}\vspace{-1cm}
\includegraphics[width=8cm]{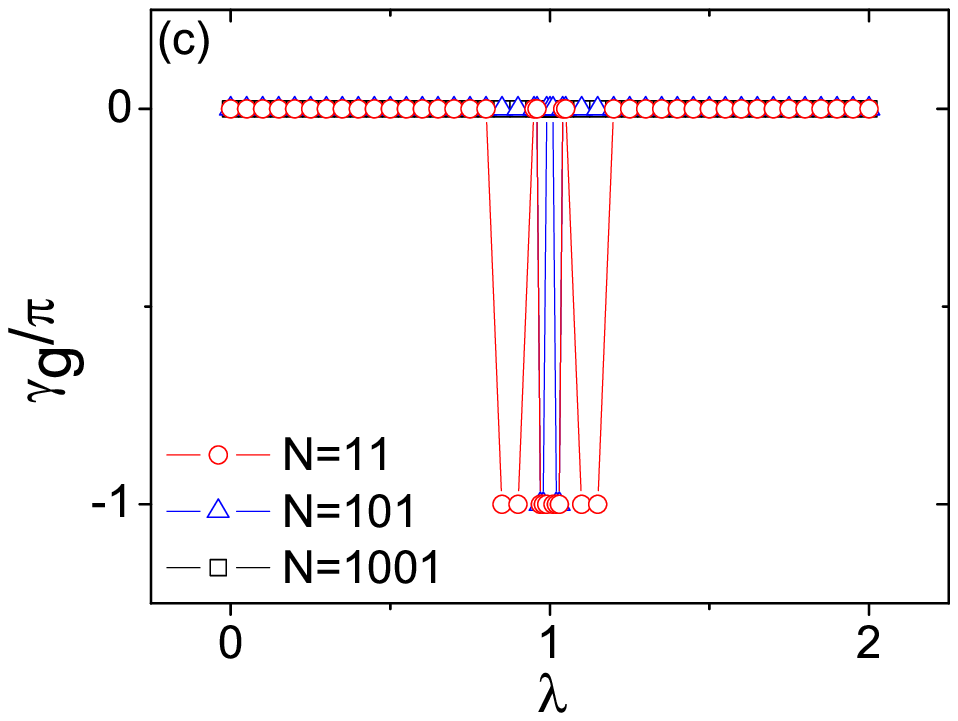}
\caption{\label{d1g} ( Color online )(a) $z$ for $\lambda=1$ (
plotted by a logarithm ) vs the particle number $N$ for $D=1$; (b)
the asymptotic behavior of $z$ closed to transition point
$\lambda=1$ for $D=1$; (c) the geometric phase $\gamma_g$ for $D=1$.
We have chosen $\gamma=1$ for all plots.}
\end{figure}

\begin{figure}
\includegraphics{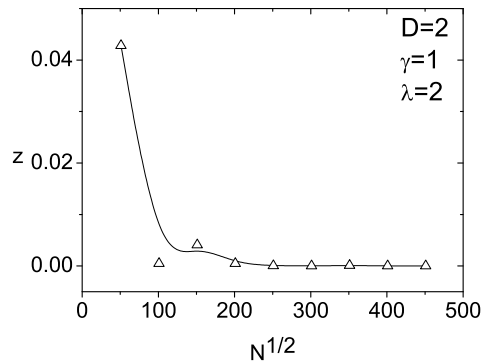}
\caption{\label{z2c} $z$ for $D=2$ vs particle number $N$ at phase
transition point $\lambda=2$. }
\end{figure}

$D=1$. It is the well-known one-dimensional $XY$ model for this
case, in which there is a quantum phase transition at $\lambda=1$
because of the degeneracy of the ground-state energy. $z$ are
plotted with $\lambda$ respectively in Fig. \ref{ff} (a). It is
obvious that $z$ has an dropping and then an abrupt increment when
one approaches the phase transition point $\lambda_c=1$. Moreover
our calculation shows that $z$ tends to be zero with
$\lambda\rightarrow1$ as shown in Fig. \ref{d1g} (b),  and at exact
$\lambda_c=1$ $z$ tends to be 1 with the increase of the lattice
site number as shown in Fig. \ref{d1g} (a) \cite{note2}. $\gamma_g$
have also been plotted in Fig. \ref{d1g} (c), in which a rapid
oscillation happens closed to $\lambda_c=1$. These phenomena mean
that $\gamma_g$ is ill-defined closed to $\lambda=1$ and has an
abrupt change at the phase transition point. Since the energy gap
vanished only at $\lambda_c=1$, the singularity of $\gamma_g$ and
$z$ would be directly related to the degeneracy of the ground state
energy. It also hints that  one could mark the transition point by
detecting the point where $z=0$.

$D=2$. With the increment of dimensionality, the situation becomes
more complex. We have plotted $z$ in Fig. \ref{ff} (b). It is
obvious that two different regions can be identified as
$\lambda\in[0,2]$ in which $z$ is disordered, and $\lambda>2$ in
which $z$ is an increasing function of $\lambda$. With respect that
the disappearance of the energy gap above the ground state happens
when $\lambda\in[0,2]$, $z$ presents a clear identification of the
phase diagram. It is a reasonable speculation from Fig. \ref{ff} (b)
that $z$ may tend to be zero under the thermodynamic limit when
$\lambda\in[0,2]$. Then $\gamma_g$ is under thermodynamic limit
\begin{eqnarray}
\gamma_g= \begin{cases} \text{undetermined}, &\lambda\le2 \\ 0,
&\lambda>2.\end{cases}
\end{eqnarray}
Our calculation also shows that $z$ tends to be zero with the
increment of $N$ at  the exact transition point $\lambda_c=2$, as
shown in Fig. \ref{z2c}. Similar to the case of $D=1$, it may be
desirable to find the point $z=0$ as a way of detecting the phase
transition.

$D=3$. This case is very similar to that of $D=2$, except the phase
transition happens at $\lambda_c=3$. $z$ has been shown in Fig.
\ref{ff} (c). However in this case $z$ seems unlikely to detect the
phase transition since the data in the figure has a smoothing
changes at the phase transition point for large $N$, as shown in the
inset of Fig. \ref{ff} (c). Only for $N=11^3$, there is an abrupt
changs of $z$ near to the phase transition point. One should note
that $z$ tends to be zero when $\lambda\in[0,3]$, in which the
energy gap above the ground state disappears.

From the discussions above, one can note the great impact of the
degeneracy of the ground-state energy on the geometric phase
$\gamma_g$ or $z$; The degeneracy of the ground-state energy leads
to $z=0$ or the ill-defined $\gamma_g$. However, this conclusion
would not be made until the next example is studied in which the
energy gap above the ground state does not disappear. It is
interesting to give a further discussion of the geometric phase in
this nontrivial case.

Unfortunately  Eq. \eqref{kff} seems unsuccessful in characterizing
the transitions for $\gamma=0$ (the tight-binding model) since in
this case $K_{\mu}$ is completely undetermined when
$\lambda-\sum_{\alpha}\cos k_{\alpha}=0$. With respect to Eqs.
\eqref{hd} and \eqref{eta}, it means that $[H, \eta]=0$ since
$[\sum_{\langle {\mathbf{i,j}} \rangle}c_{\mathbf{i}}^\dagger
c_{\mathbf{j}}, \sum_{\mathbf{i}} c_{\mathbf{i}}^\dagger
c_{\mathbf{i}}]=0$ in this special case. Hence our discussion
excludes this special situation since one has trivial results. One
should note that the phase of $\lambda=0$ with $\gamma>0$ in $D=2$
also cannot be identified by $z$ since the transition comes from the
deformation of the Fermi surface instead of the degeneracy of the
ground state.

\subsection{Su-Schrieffer-Heeger (SSH) model}
\begin{figure}
\includegraphics[width=8cm]{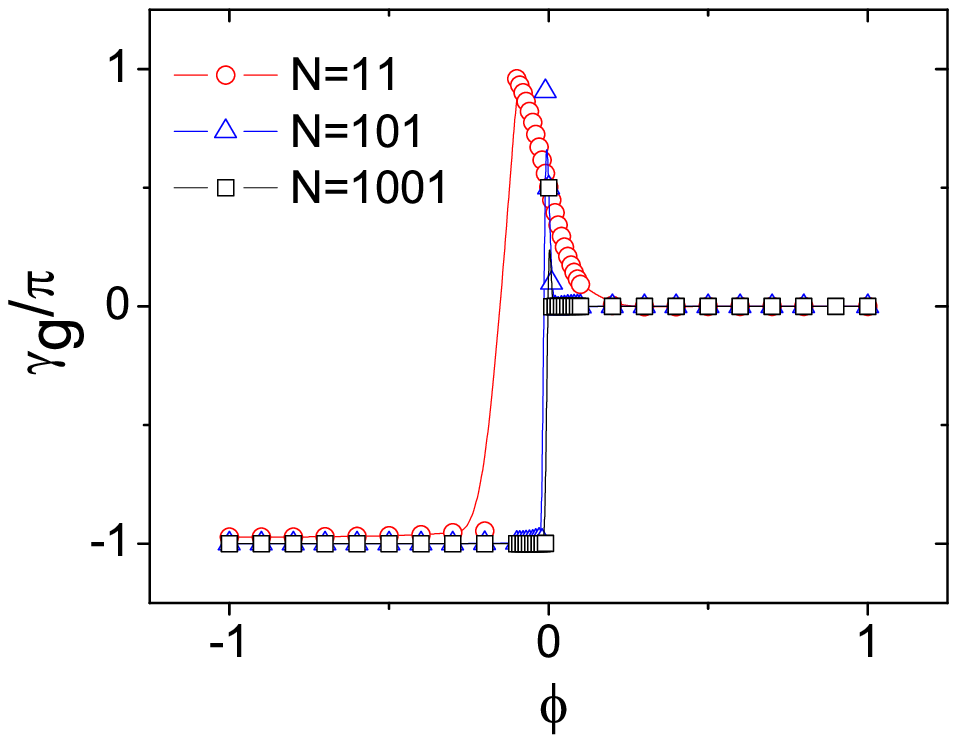}
\includegraphics[width=8cm]{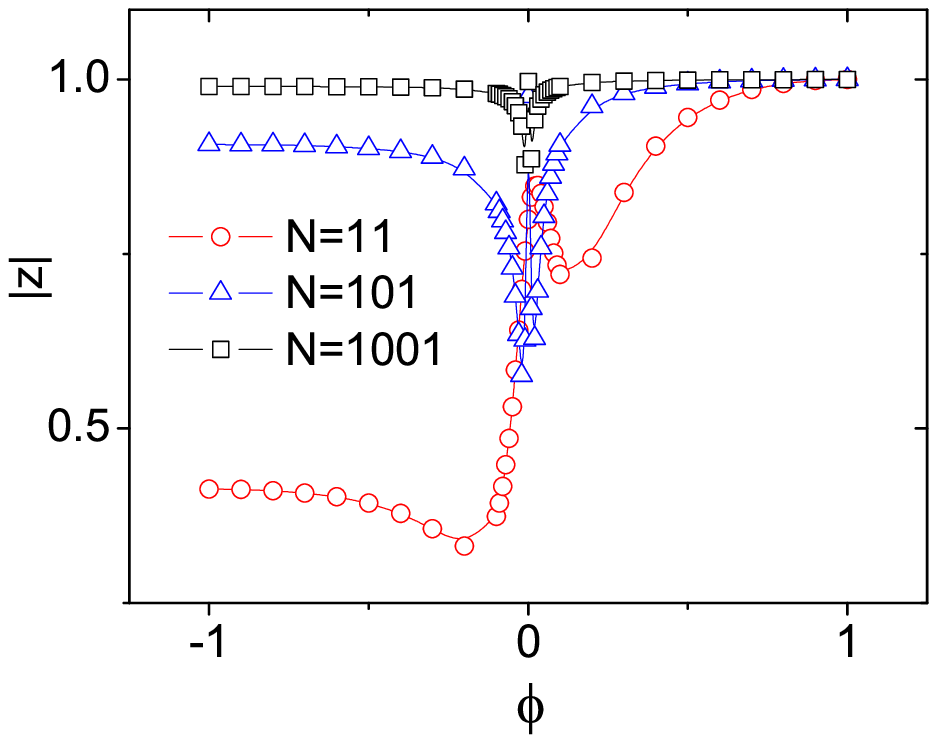}
\caption{\label{sshz} ( Color online ) $\gamma_g$ and $|z|$ for the
SSH model vs. the parameter $\phi$.}
\end{figure}
Another example is the Su-Schrieffer-Heeger (SSH) model, which is
also exactly solvable. The 1D tight-binding Hamiltonian for the SSH
model for a chain of polyacetylene is given by \cite{heeger}
\begin{equation}\label{ssh}
H=\sum_{l=1}^{N}t(-1 + (-1)^l\phi_l)(c_l^{\dagger}c_{l+1}+ h.c.),
\end{equation}
in which $\phi_l$ represents the dimerization at the $l$th site and
an alternating sign of the hopping elements reflects dimerization
between the carbon atoms in the molecule. Without the loss of
generality, it is convenient to neglect the kinetic energy in the
system and take $\phi_l=\phi$, $t=1$ \cite{ryu2}. There is a
critical point at $\phi=0$ ,which divides the ground states into two
different types. One should point out that SSH has a gaped
excitation for any $\phi\in[-1,1]$ and the quantum phase transition
comes from the excitation of the boundary states\cite{ryu2}.

One can find from the following calculation that $\gamma_g$ can
discriminate the two phases and the boundary between them. Defining
$c_x=(c_l, c_{l+1})^T$, the Hamiltonian becomes
\begin{eqnarray}
H&=&\sum_{x=1}^{N}c_x^{\dagger}\left(\begin{array}{cc}0&-(1+\phi)\\-(1+\phi)&0\end{array}\right)c_x\nonumber\\
&+&\left
[c_x^{\dagger}\left(\begin{array}{cc}0&0\\-1+\phi&0\end{array}\right)c_{x+1}+h.c.\right
]
\end{eqnarray}
Imposing the periodic boundary condition and Fourier transformation,
we then have
\begin{equation}
\mathscr{H}(k)=
\left(\begin{array}{cc}0&-(1+\phi)-(1-\phi)e^{-ik}\\-(1+\phi)-(1-\phi)e^{ik}&0\end{array}\right),
\end{equation}
It is obvious that $\mathbf{R}(k)=(-(1+\phi)-(1-\phi)\cos k,
-(1-\phi)\sin k, 0)$ and $R_{y}(k)=- R_{y}(-k)$ is satisfied. Then
\begin{equation}
z=\prod_k (\cos\frac{2\sqrt{2}\pi}{N}|K'| -
\frac{i\text{Im}[K']}{\sqrt{2}|K'|}\sin\frac{2\sqrt{2}\pi}{N}|K'|),
\end{equation}
in which $K'=-\frac{i}{2}\frac{(1-\phi)^2+(1-\phi^2)\cos
k}{(1-\phi)^2\sin^2k+[1+\phi+(1-\phi)\cos k]^2}$.

We plot $\gamma_g$ against $\phi$ with different site numbers in
Fig. \ref{sshz}. It is obvious that $\gamma_g$ is $- \pi$ for
$\phi\in[-1, 0)$ and zero for $\phi\in (0, 1]$ from Fig. \ref{sshz}.
Moreover $\gamma_g$ tends to be $\pi$ at exact phase transition
point $\phi=0$. Furthermore our calculation shows that $|z|$ is not
zero, which means that one cannot detect the phase transition by
finding the point $z=0$. Since the energy gap for the ground state
is nonvanishing in this model, the geometric phase $\gamma_g$ can be
well-defined for any $\phi$. With respect to the discussion for the
$D$-dimensional free-fermion system, it is evident that $\gamma_g$
or $z$ are directly related to the degeneracy of the ground state.

\section{discussions and conclusions}
Given the two examples, some comments should be presented in this
section. In this paper the twist operator Eq. \eqref{tw} has been
introduced and its ground-state expectation $z$ has been calculated
to define the geometric phase Eq. \eqref{gp} for the two-band model.
Although the absent of general results, the exact expression of $z$
can be obtained in a special case, which provides the ability to
detect the details of the geometric phase adjacent to the phase
transition points. With respect to the discussions for two
representative examples-D-dimensional free-fermions model and the
Su-Schrieffer-Heeger model, some distinguished properties of
$\gamma_g$ or $z$ have been found in our calculations.

First when the degeneracy of the ground state happens, $z$ tends to
be zero and the geometric phase $\gamma_g$ is ill-defined in this
case, as shown in Figs. \ref{ff}, while $\gamma_g$ can be
well-defined when the energy of the ground state is nondegenerate,
as shown in Fig. \ref{sshz}. This phenomenon clearly displays the
intimation connection between geometric phase, defined by the
ground-state expectation value of the twist operator, and the
degeneracy of the ground state in many-body systems. Consequently
one can find the nodal structure of the geometric phase (the
situation that the geometric phase is ill-defined because of $z=0$)
to detect the phase transition originated from the degeneracy of the
ground state. The nodal structure of geometric phase is introduced
by Fillip and Sj\"oqvist for the description of experimental measure
of the geometric phase based on the interference, which
characterizes the condition for the disappearance of the fringes and
then the geometric phase is ill-defined. Second geometric phase can
also present the phase diagram even if there is a energy gap above
the ground state, as shown in Fig. \ref{sshz}, in which two
different phases defined by $\gamma_g=-\pi$ and $0$ can be
identified and the phase transition point is marked by the
discontinued variation of geometric phase. In a word the geometric
phase $\gamma_g$ displays the ability to mark the phase diagram in
this discussion, whether the phase is determined by the degeneracy
of the ground state or not. Hence the geometric phase can provide a
more popular depiction for the phase transition.

However there still exist some problems. First is that the geometric
phase seems to fail to characterize the tight-bond model ($\gamma=0$
in Eq. \eqref{hd}). It is the reason that $[H, \eta]=0$, and the
twist operator has a trivial effect on the ground state.  Second the
geometric phase fails to detect some phase transitions not
originated from the degeneracy of the ground state, such as the
transition from the deformation of the Fermi surface. Thirdly
$\gamma_g$ or $z$ seems unable to detect the broken of symmetry
which happens in the 1D spin-1/2 $XY$ model, as shown in Fig.
\ref{ff} (a). Although there exists some defects, the geometric
phase defined by the twist operator provides one another way of
detecting the phase diagram for many-body systems. Moreover the
flexibility of choosing $n_x$ in the definition of the twist
operator implies that one could properly choose different physical
quantities $n_x$ for the description of different properties of the
system.

\emph{Note added}. Recently  we become aware of  a paper which also
focuses on the connection between the geometric phase and the
quantum phase transition by numerical evaluation \cite{hkh}. In this
paper, three different phases in gapped spin chains can be defined
by the geometric phase $\gamma=0, \pi,$ and undefined respectively,
which is similar to our conclusions.

\emph{Acknowledgement} The author(H.T.C.) acknowledges the support
of NSF of China, Grant No. 10747195.


\begin{thebibliography}{99}
\bibitem{footnote} It is convenient for a two-levle system to write
the desity matrix as $\rho=1/2(1+ \mathbf{R}\cdot\mathbf{\sigma})$,
in which the vector $\mathbf{R}$ is real three-dimensional vector
and $|\mathbf{R}|\leq 1$. This vectoer is well known as the Bloch
vector.
\bibitem{note}Since Eq. \eqref{hd} is spatially isotropic, it has no
difference for $z_{\mu} (\mu=1\dots D)$ with $D>1$ and these
plottings are chosen for $z_1$.
\bibitem{note2}The indetermination of $K_{\mu}$ when $\lambda=1,
k_{\alpha}=0$ can be resolvable by L'Hospital rule. Our calculation
shows that $K_{\mu}$ is 0 for this case. For $D=2,3$, the similar
calculations are applied.
\end{thebibliography}
\end{document}